\documentclass[preprint,authoryear,12pt]{elsarticle-prepr}

\usepackage{color}

\definecolor{mrk}{RGB}{160,0,0}

\usepackage{graphics}
\usepackage{graphicx}
\usepackage{amssymb}

\usepackage[ps2pdf,%
a4paper=true,%
breaklinks=true,%
colorlinks=true,%
pdfauthor={Gordovskyy et al.},%
pdftitle={Mini-review on particle acceleration in solar flares}%
]{hyperref}

\journal{ }

\begin{document}

\begin{frontmatter}

\title{Combining MHD and kinetic modelling of solar flares}

\author{Mykola Gordovskyy\corref{cor}}
\cortext[cor]{Corresponding author}
\ead{mykola.gordovskyy@manchester.ac.uk}
\author{Philippa Browning}
\address{Jodrell Bank Centre for Astrophysics, School of Physics and Astronomy, University of Manchester, Manchester M13 9PL, UK}

\author{Rui F. Pinto}
\address{Institut de Recherche en Astrophysique et Plan\'{e}tologie, Universit\'{e} de Toulouse, OMP/CNRS, 31028 Toulouse Cedex 4, France}

\begin{abstract}
Solar flares are explosive events in the solar corona, representing fast conversion of magnetic energy into thermal and  kinetic energy, and hence radiation, due to magnetic reconnection. Modelling is essential for understanding and predicting these events. However, self-consistent modelling is extremely difficult due to the vast spatial and temporal scale separation between processes involving thermal plasma (normally considered using magnetohydrodynamic (MHD) approach) and non-thermal plasma (requiring a kinetic approach). In this mini-review we consider different approaches aimed at bridging the gap between fluid and kinetic modelling of solar flares. Two types of approaches are discussed: combined MHD/test-particle (MHDTP) models, which can be used for modelling the flaring corona with relatively small numbers of energetic particles, and hybrid fluid-kinetic methods, which can be used for modelling stronger events with higher numbers of energetic particles. Two specific examples are discussed in more detail: MHDTP models of magnetic reconnection and particle acceleration in kink-unstable twisted coronal loops, and a novel reduced-kinetic model of particle transport.
\end{abstract}


\end{frontmatter}

\parindent=0.5 cm

\section{Introduction}

Solar flares are the most powerful non-stationary phenomena in the Solar System, releasing up to 10$^{25}$~J within 10-100~s. Magnetic reconnection in high-temperature turbulent current sheets (CS) is widely believed to be the key physical mechanism behind the flares, converting magnetic energy stored in the corona into thermal and kinetic energy, and subsequently  radiation in many wavelengths of the electromagnetic spectrum. A substantial part of the flare energy is carried  by high-energy non-thermal particles: electrons with energies typically from around 10 keV to 100 MeV, and also  ions with energies up to hundreds of MeV \citep[see e.g.][for review]{hury95,zhae11,vilm12,benz17,klda17}.  Electrons in the flaring corona are detected observationally both through bremsstrahlung as they impinge in on the dense chromosphere, and sometimes in the corona itself, as well as through radio from gyrosychrotron and plasma emission. Some energetic particles from flares may also escape into the heliosphere as "solar energetic particles" where they can be detected  {\it in situ}.  Often, solar flares are associated with Coronal Mass Ejections (CME) and other eruptions affecting the outer corona and heliosphere, and causing major disturbances in the near-Earth space environment,  with solar energetic particles also playing an essential role.

As one of the main manifestations of solar activity, flares are essential for understanding both the nature of the magnetic Sun and the physics of many high-energy astrophysical phenomena. Being one of the key factors behind  space weather, flares are also of practical importance \citep[see][]{schw06,chen11,saie13}. Hence, it is essential to have a comprehensive, self-consistent model of solar flares, accounting both for thermal and non-thermal plasma. One of the major unsolved problems is to explain the process by which energetic particles are accelerated, and a number of mechanisms are proposed, including  super-Dreicer electric fields in the reconnecting CS, second-order Fermi acceleration through turbulence, shocks or waves \citep[e.g.][]{asch02,zhae11,vilm12,benz17}. Collapsing magnetic traps may also play a role, but are unlikely to be the primary source of acceleration \citep{some02,grne09}. The problem of flare particle acceleration should be addressed in the wider context of understanding the origin of energetic charged particles in a wide range of laboratory, space and astrophysical plasmas \citep[see e.g.][]{mctu17}.

Scale separation is the main obstacle for self-consistent flare modelling. The  overall size of the flaring region is $10^7-10^8$~m, and evolution of electromagnetic fields and thermal plasma at these large scales can be described in terms of magnetohydrodynamics (MHD). However,  particle trajectories have spatial scales as small as $\sim10^{-2}$~m (typical electron Larmor radii) and the  typical thickness of a CS is likely to be  $100-1000$~m. Similarly, global time-scales of flare evolution are of the order of minutes or hours, while kinetic processes have time-scales as small as $10^{-9}$~s (inverse plasma frequency, electron cyclotron frequency). Hence, processes such as wave-particle interaction at small-scales and evolution of non-thermal plasma (energetic particles) in the flaring corona cannot be described using fluid approach ({\it i.e.} MHD) and require a kinetic description. That is why magnetic reconnection, particle acceleration, and energetic particle transport are usually considered separately. However, it is now clear that wave-particle interaction at small scales is necessary for fast magnetic reconnection, and  plasma within and around the  CS is  essentially non-thermal.  Hence, modelling the primary energy release in solar flares must take  account of plasma kinetics, and cannot be treated by a fluid model.

A fully-kinetic description ({\it i.e.} using Maxwell-Vlasov or similar formalism) would be the most obvious  means of  solar flare modelling. However, taking into account the need to resolve very small scales, such as electron Larmor radii ($\sim10^{-2}$~m), this method is extremely computationally expensive and the  physical size of the model is restricted to less than $10^2-10^4$~m, depending on the model geometry. The use of the guiding-centre (or drift-kinetic) approximation removes the need to resolve particle gyro-radii;  however, it would be still necessary to resolve electrostatic scales ({\i.e.} Debye length,  $\sim10^{-2}$~m, and inverse plasma frequency, $<10^{-8}$~s). That is why fully kinetic methods, usually implemented using the particle-in-cell aproach, are normally exploited to develop models with rather unrealistic parameters. Still, these models make it possible to study some fundamental properties of  magnetic reconnection in hot solar and space  plasmas:  in particular,  to see how kinetic effects give rise to anomalous transport effects, which are necessary for fast magnetic reconnection, and to investigate particle acceleration and thermal plasma response to energetic particle transport \citep[e.g.][]{taje87,hewe88,mcce90,drae06,tsha08,sizh09,baue13,guoe14,lie14,grts16}.

Solar plasma is expected to remain quasi-neutral at all times at spatial scales larger than Debye length. Hence, it is appropriate to consider a Vlasov model  with zero net-electric-charge density everywhere. In this case, a low-frequency limit of Maxwell equations can be considered: essentially, ignoring electrostatic effects \citep{trca15}. This would make it possible to drastically increase the size of the model domain, since only electromagnetic scales would need to be resolved ($1-100$~m and $10^{-5}-10^{-4}$~s). Still, taking into account that a kinetic model has a larger number of dimensions compared to an MHD model, due to the velocity dimensions of phase space, even quasi-neutral Maxwell-Vlasov models are computationally expensive.

An alternative to fully kinetic treatments  would be a combination of MHD describing the thermal plasma and electromagnetic field, and a kinetic method describing the non-thermal plasma. The goal of this paper is to provide a critical review of such approaches, in the context of modelling solar flares, considering both past work and potential future directions. In Section~2 we consider the approach based on the combination of an MHD and test-particle models, providing an overview of how this approach has been used to study  particle acceleration by magnetic reconnection in solar flares as well as focusing on the  specific example of modelling thermal and nonthermal plasma in a kink-unstable twisted coronal loop. In Section~3 we discuss various hybrid schemes with MHD and kinetic methods working within the same computational model, and outline a novel "reduced kinetics" approach developed to study particle transport - and potentially, magnetic reconnection - in solar flares.

\section{Combined MHD/test-particle modelling}

\subsection{MHD/test-particle approach}

Perhaps, the most obvious approach to fluid-kinetic modelling of the coronal plasma is the combination of the MHD and test-particle approaches.  This involves solving the equations of motion for ions and electrons in given electromagnetic fields arising from an MHD model; any electromagnetic fields generated by the test-particles are neglected. This approximation is valid so long 
as the number of non-thermal particles is small compared with the background thermal plasma.  The background fields may be specified through an analytical solution to the MHD equations, or some simple idealised model representative of MHD behaviour, or from numerical solution of the MHD equations. Depending on the field  configuration, the particle equations of motion may be solved 
either using the full Lorentz equation of motion, or the guiding-centre approximation - the latter is generally relevant in the case of strongly-magnetised plasmas in the solar corona, except close to magnetic nulls. A more powerful tool, particularly for fields containing null points, is a code which switches between guiding-centre and full-trajectory calculations, dependent on the local particle gyro-radii and field gradients \citep{broe10,stan13}. A further modelling choice is whether to use the relativistic equation of motion, bearing in mind that relativistic effects are particularly relevant for high-energy flare electrons. Furthermore, although majority of test-particle models ignore collisions with the background plasma, some test-particle simulations include Coulomb collisions and scattering due to micro-turbulence \citep[e.g.][]{hame05,kaba06,gore14,bure14,bore17}.

\begin{figure*}   
\centerline{\includegraphics[width=0.83\textwidth,clip=]{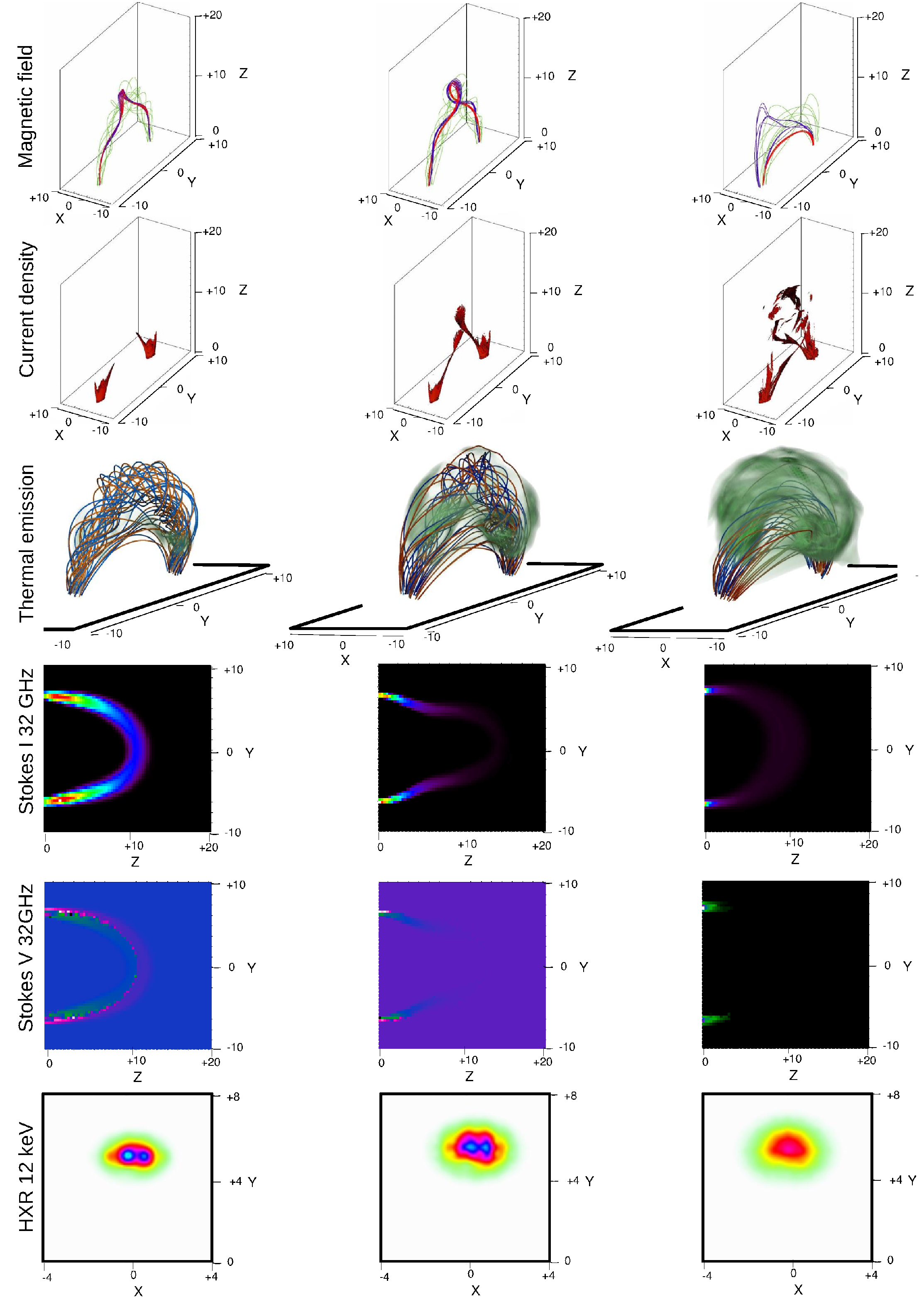}}
\caption{\scriptsize Model of magnetic reconnection in twisted loop in the solar corona. Panels in left, middle and right columns correspond to onset of magnetic reconnection, stage with fastest energy release and decay of the reconnection event, respectively. Please note, that axes orientation is different in different row. First row shows selected magnetic field lines (different colours are used to show change of magnetic connectivity), second row shows surfaces of equal current density \citep{gore14}, third row shows thermal emission in soft X-ray continuun arount 2~keV. These panels show the full model domain in axonometric projections. Fourth and fifth rows show microwave intensity and microwave polarisation maps. These panels show emission from the whole model domain mapped in Y-Z plane (i.e., the loop is observed from the side. Finally, the sixth row shows hard X-ray intensity map of one of the foot-points. These panels show emission approximately from a half of the domain mapped in X-Y plane (i.e., the loop is observed from the top.)}
\end{figure*}
 
There are numerous implementations of test-particle models, which  can be characterised using two key features, as follows. Firstly, different implementations use either time- dependent or quasi-stationary electromagnetic fields. In principle, MHD time-scales can be much longer than characteristic kinetic time-scales for some particles, validating the latter approach. For instance,  with length scales of 10$^7$~m and magnetic field of about 100~G, the Alfv\'{e}n time-scale is about 1-10~s, while the  electron gyro-period is about 10$^{-8}$~s, and typical electron acceleration time is 0.1-1~s. Hence, the assumtion of stationary electromagnetic fields would not affect trajectories of individual test electrons. However, the total energy aquired by a population of test particles may be strongly dependent on the evolution of electromagnetic fields; for instance, the total energy gain would increase with the duration of a transient reconnection event. Furthermore, protons have much longer transit times, and the electromagnetic fields are likely to evolve even over the trajectory of an in individual proton. 
Secondly, many implementations use initial test-particle distribution which is not statistically representative of plasma in the MHD model. By ``statistically-representative'' we mean test-particle populations corresponding to the distribution function 
\[
f(\vec{r},v) \sim \rho(\vec{r}) v^2 \exp \left( - \frac{m v^2}{2 k_B T(\vec{r})} \right),
\]
where $\rho(\vec{r})$ and $T(\vec{r})$ are density and temperature distribution in the MHD model, and $v$ are the particle velocities in the plasma frame of reference, with isotropic pitch-angle distribution. The models with either stationary electro-magnetic field or statistically non-representative initial test-particle populations are useful tools to investigate individual particle trajectories. However, as with any Monte Carlo-type approach, they might not be able accurately  to reproduce characteristics of the total particle populations, such as the spatial distribution of energetic particles or their energy spectra.

Because this method is partially based on resistive MHD, one of its important drawbacks is a need for anomalous resistivity $\eta$, which is, generally, unknown. This parameter is important not only for field evolution but also for particles, as it determines the value of the parallel electric field, so that particle energies, to a first approximation, are proportional to $\eta$. (In reality, it is more complicated, because $\eta$ also affects the size and lifetime of the accelerating region.) There are different approaches to setting the anomalous resistivity, from uniform and constant $\eta$ to $\eta(\vec{r},t)$ determined by local current density and other plasma parameters, so that the spatial distribution of anomalous resistivity ``mimics'' the effect of ion-acoustic instability, which is often considered to be a trigger for anomalous resistivity in the corona \cite[see discussion in][]{bare11}. The amplitude of anomalous resistivity determines the duration of a transient magnetic reconnection event. Hence, the typical duration of the impulsive phase can be used to constrain the resistivity amplitude, at least by the order of magnetitude \cite[see e.g.][]{gore14}.

\subsection{MHD/test-particle models of particle acceleration in solar corona}

The MHD/test-particle (MHDTP) approach has been extensively used to reveal the role of different  mechanisms for particle acceleration and the effect of the field structure on particle acceleration in flares. A major focus has been the investigation of electromagnetic fields representative of magnetic reconnection. A large body of work considers test-particles in idealised analytical models for quasi-stationary configurations containing a current sheet (field reversal) or an X-point, both with and without an out-of-plane magnetic field component or ``guide field'', building on the pioneering work of \citet{spei65} and several other analytical studies \cite[e.g.][]{liso93,litv96}. For example, \citet{saka90} calculated trajectories of protons and electrons in a quasi-stationary field configuration corresponding to two interacting magnetic fluxtubes. \citet{klie94} developed a 2D test-particle model with quasti-stationary fields and studied charged particle trajectories around X- and O-points, demonstrating importance of the convective electric field for particle acceleration near magnetic islands. Later \citet{vebr97} and \citet{brve01} considered a 2D model with X-point magnetic fields, showing the utility of the guiding-centre approach for test-particle modelling, and revealing the importance of non-unifomity in the electric drift in particle acceleration as well as acceleration by the parallel electric field. Investigations of test-particles in 1D current layers  revealed possibility of electron-ion separation in presence of strong guide-fields \citep{zhgo04,zhgo05}, which has been later confirmed using more realistic 2D X-point model \citep{wone05}. Using a 2D linearised MHD and a test-particle code accounting for Coulomb collisions, \citep{hame05} further investigated the effect of electron-ion separation in the model of magnetic reconnection around x-point, as well as showed the effect of collisions on the resulting energy spectra of X-ray producing electrons. Also, \citet{hafl06} applying the test-particle approach to a simple current sheets configurations showed how X-point parameters affect resulting power-law tails of particle spectra.

Building on the studies considering idealised, quasi-stationary configurations, more complicated, realistic MHD/test-particle models utilise time-dependent electromagnetic fields arising from MHD simulations. Perhaps, one of the first computational studies of this kind has been performed by \citet{sate82}, revealing proton and electron behaviour during magnetic reconnection in an externally-disturbed neutral current sheet. In more recent years, various MHD/test-particle studies with evolving fields have been used to study particle acceleration in 2D and 3D configurations. Thus, \citet{bire04} considered particle acceleration in the Earth magnetotail using MHDTP approach demonstrating relative importance of different acceleration mechanisms. Later \citet{gore10a,gore10b} developed a model coupling test-particles to time-evolving MHD fields in order to  investigate particle acceleration in current sheets during forced magnetic reconnection \citep{haku85}. The main mechanism accelerating particles in these models is DC parallel electric field, with rather negligible contribution of the betatron mechanism due to magnetic islands contraction. The study revealed the formation of two energetic particle populations: particles travelling in the ``open'' field and particles trapped by contracting magnetic islands, and also  demonstrated that particles can be accelerated abruptly, with energy increasing in a jump-like manner when a particle goes through a thin electric field layer. The effect of magnetic islands on particle acceleration has been further investigated using MHDTP by \citet{lie17}. They also demonstrate formation of two particle populations, although trapped particles in their model have softer power-law energy spectra compared to the particles in open fields. Later, \citet{zhoe15,zhoe16} used the MHDTP approach in order to investigate the role of different effects and different scales on electron acceleration in 2D reconnecting current sheets. This study is based on the MHD model by \citet{bare10, bare11}, which uses the automatic mesh refinement algorithm, allowing to study smaller spatial scales. 

Another mechanism interesting in the context of solar flares which can be investigated using 2.5D MHDTP method is particle acceleration in collapsing magnetic traps \citep{soko97}. Generally, large-scale magnetic field relaxation should result in effective ``shortening'' of field lines, producing some acceleration - as occurs in the Earth's magnetotail \citep{bire17}. Viability of this effect has been tested using several models involving test-particle simulations \citep[e.g.][]{giue05,kaba06,grne09,grae12,bire17}. It appears that in realistic solar flare conditions this mechanism can accelerate particles to very moderate energies of only few keV. However, the electric field required to produce runaway electrons depends on the particle initial energy as $1/\mathcal{E}_{ini}$. Therefore, pre-acceleration in collapsing magnetic traps may be very important in two-stage acceleration scenarios, as well as in scenarios including local acceleration and re-acceleration \cite[see e.g.][]{broe09}.

The coronal magnetic field is naturally complex and in reality lacks the symmetries imposed by 2D models.  Considering test-particles in 2D fields can over-estimate acceleration, since particles may move arbitrarily large distances along the electric field in the invariant direction. Furthermore, the nature of 3D magnetic reconnection differs in significant ways from traditional 2D models \citep[e.g.][]{pont11}.  \citet{dabr05} were the first to point out the potential importance of 3D magnetic null points as sites of particles acceleration, and  many observational studies have confrimed the presence of energetic particles in flares associiated with reconnecting 3D nulls \citep[e.g.][]{mase09}. A body of subsequent work using the  test-particle approach in  3D MHD configurations has  explored electron and ion behaviour in the vicinity of null-points and separators. \citet{dabr06,dabr08} and \citet{broe10} investigated trajectories of particles accelerated close to the null-point in a 3D quasi-stationary fan reconnection configuration. They compared particle acceleration efficiency in fan and spine configurations, as well as demonstrated  the formation of two populations -- escaping and trapped energetic particles. Later, \citet{guoe10} investigated particle behaviour around 3D null-point using time-dependent MHD simulations and a full trajectory test-particle model. They particularly focused on the role of convective magnetic field, concluding that it should be more efficient in accelerating protons compared to electrons. Also, they discussed limitations of this approach due to finite spatial resolution. Particle acceleration in analytical electromagnetic fields which are exact solutions of the 3D MHD equations was investigated by \citet{stbr12}, showing that the fan reconnection regime is more effective for particle acceleration than spine reconnection, and that particles are mainly accelerated outside the current sheet due to non-uniform drifts. Combined MHDTP model of particle acceleration in the vicinity of magnetic separator developed by \citet{thre15} based on time-dependent kinematic model managed to produce particle energy spectra similar to the power-law distributions inferred from HXR observations, although,  as  in the majority of test-particle simulations, the energy spectra appear to be very hard, with spectral indices between 1.0-2.0.

Alternative to these scenarios with regular and localised  electric fields, are models with strongly fragmented or chaotic electric field distributions. These models are particularly interesting because they usually have energy release distributed within larger volumes, thus, potentially producing much larger numbers of energetic particles, compared to the standard reconnecting current sheet models \citep{vlae09,care12}.  Studies of test-particles in turbulent magnetic fields show how particles are accelerated both due to localised parallel electric fields and due to second-order Fermi acceleration \citep{arvl04,dmie04,isle17}. A test-particle study of a stationary 3D snapshot of stressed coronal field demonstrated how this configuration can result in bulk production of energetic particles \citep{ture05, ture06, care06}. Later, \citet{gobr11,gobr12} used test-particle simulations to investigate electron and ion energisation in twisted coronal loops, in which the kink instability triggers the formation of fragmented current structures, discussed in more detail below. Interactions and mergers of adjacent twisted flux ropes can release substantial amounts of free magnetic energy through magnetic reconnection, which can provide efficient particle acceleration. This may occur through the ideal tilt instability if the flux ropes carry opposing currents, driving fast reconnection and generating power-law tails of non-thermal particles, with slopes dependent on the resistivity profile \citep{ripp17a,ripp17b}. Twisted flux ropes carrying like currents may interact through an MHD avalanche, in which kink instability in one flux rope may lead to merger with neighbouring stable flux ropes and release of magnetic energy from the stable flux ropes \citep{hooe16}. Test-particle simulations coupled with 3D  MHD simulations of this scenario   \citep{thre18}  reveal two phases of particle acceleration, firstly due to electric fields associated with the fragmented current sheets in the initial kink instability, and a second phase as the loops reconnect and merge. Reconnections between different magnetic flux systems may also occur during flux emergence, and particle acceleration in this situation has been studied, using test-particles and 3D MHD simulations, by \citet{roga10}.

\subsection{MHDTP model of reconnecting twisted coronal loop}

Let us use the  kink-unstable twisted coronal loop configuration  for a more detailed demonstration of the MHDTP approach \citep[see][and references below]{gobr11}. The scenario is based on several analytical and computational studies \citep[including][]{brva03, broe08,hooe09}, which showed that the onset of ideal kink instability in  twisted coronal loops leads to the formation of fragmented current sheets throughout the loop volume, in which magnetic reconnection releases magnetic energy.  These initial studies  considered a cylindrical model loop whch is initially in a kink-unstable equilibrium state, but more recent work has extended this to more realistic geometries, including curved coronal loops with varying degrees of field convergence \citep{bare16}. The more recent  MHDTP model combines 3D MHD simulations of magnetic reconnection in a twisted loop in a gravitationally-stratified atmosphere with Lare3D MHD code \citep{arbe01} with the model of particle acceleration and transport  using the GCA test-particle code \citep{gore10b,gobr11}. The latter is based on relativistic guiding-centre approximation; the code utilises time-dependent electric and magnetic fields, and plasma density defined on a discrete grid,  and calculates field and density values for individual particle positions using three- or four- dimensional linear interpolation. Thus, the electromagnetic fields for the particle trajectories evolve in time in accordance with the MHD dynamics. Also, the code can account for particle scattering and energy losses due to Coulomb collisions, similarly to \citep{hame05,kaba06}. 

Slowly-varying parallel electric field is the main particle acceleration mechanism in this model, although the loop contraction is likely to make small contribution as well. The typical acceleration efficiency is about 0.04-0.07 \citep{gobr12},  validitating the test-particle approach for this scenario. The evolution of the loop, as well as the  thermal and non-thermal emission produced by the loop,  is shown in Figure~1.

This model shows that energy release in reconnecting twisted loops is distributed through a large  volume ($10^{19}-10^{20}$~m$^3$, volume of a coronal loop), compared with  energy release localised in small volume of a current layer (approximately $10^{14}-10^{18}$~m$^3$) expected from the ``standard'' flare model \citep{shie95}. This makes twisted loop configurations a good alternative to the standard model, as it can, potentially, help solving the so-called particle number problem \citep{broe09}, as well as reduce particle energy losses during transport (due to the return current and various scattering effects). Whilst this configuration is not expected to explain the  majority of flares, the twisted loop scenario is a good candidate for explaining events such as isolated, simgle-loop flares \citep{asce09}, failed eruptions \citep{alee06} and others. Furthermore, twisted coronal loops are important also because they are considered one of the main elements of the phenomenological models of CME eruptions \citep[e.g.][]{gibe06}. 

A significant  practical value of this model is that it can predict  observables in different parts of electromagnetic spectrum, making it possible to distinguish solar flares and similar events in the corona caused by kink-instabilities in twisted fields from energy release in other configurations. Furthermore, the modelling approach described here could easily be adapted to alternative configurations, including multiple interacting flux ropes and the standard flare model.  Several characteristic features predicted by these models can be used for observational detection of flaring twisted loops. Thus, test-particle simulations and calculations of synthetic bremsstrahlung hard X-ray (HXR) emission showed that HXR sources produced by twisted loops should expand with time due to radial expansion of the reconnecting loop \citep{gobr11,pine16}. Typically, the size of the model HXR source increases from about 1.5~Mm to 2.5~Mm within the order of  100~s. Gradual expansion of HXR sources has been detected using RHESSI observations \citep{kone11}, although it is difficult to say what is the exact reason of the source expansion, because this effect can also be explained by strong turbulent scattering of HXR-producing electrons \cite[e.g.][]{kone14}.

Another feature related to non-thermal plasma predicted by this model is the so-called cross-loop polarisation gradient. Using the GX microwave code \citep{flku10,nite15} we calculated microwave emission produced by our MHDTP model and, similarly to earlier calculations based on analytical stationary loop models \citep{shku16}, we demonstrated that strong magnetic twist should produce a characteristic pattern of Stokes V distribution \citep{gore17}. It has been shown that this pattern can be observable even without non-thermal electrons provided the loop plasma is sufficiently hot. Also, it has been noted that if the gyrosynchrotron emission is optically thick, the pattern will be more complicated, although still detectable. This approach can also be used to study variations of radio-emission due to various MHD oscillations of the coronal loop \citep[see][and references therein]{name09}. Currently, the developed model is used to investigate pulsation of microwave emission from kink-unstable coronal loop, using the approach similar to \cite{kuze15} but with the field, thermal and non-thermal plasma parameters taken from MHDTP simulations.

Twist visible in thermal EUV and SXR emission is expected to be one of the main observational indications of magnetic field twist. That is why thermal emission from reconnecting twisted loops has been studied by combining MHD simulations with radiative simulations of thermal continuum emission \citep[e.g.][]{bote12,pine15, snoe17}. Investigation of thermal continuum emission produced by the reconnecting twisted loops explains why the observed twist is usually quite weak, with twist angle often observed to be smaller than critical angle required for kink-instability. Thus, MHD models show that instability normally occurs at twist angles 2.5$\pi$--6$\pi$ \citep{bare16}. However, twist-angle  values inferred from EUV observations are usually much lower, 2$\pi$ or less. Our model provides a very simple explanation for this: thermal emission is normally observed towards the end of the reconnection event, when plasma is very hot. By this time, the twist is substantially reduced due to magnetic reconnection \citep{pine16}. Interestingly, this also results in the Neupert-like effect: the time derivative of thermal emission intensity correlates with  non-thermal HXR intensity, {\it i.e.} thermal emission lightcurve show time delay compared to non-thermal emission lightcurve, and is more prolonged in duration,  with the the predicting timings being consistent with those  observed in  solar flares \citep[e.g.][]{tomc99}. 

Analysis of plasma turbulence in this model reveals two other interesting features \citep{gore16}. Firstly, the turbulent energy motions carry a similar amount of energy as regular plasma flows, about 10$^{-2}$ of the energy released in flares. Secondly, the amplitude of macro-turbulence (more specifically, line-of-sight velocity dispersion, the value, which determines non-thermal broadening of coronal EUV lines) correlates with the plasma temperature. This feature is in both qualitative and quantitative agreement with observations \citep{dose07,dose08}. The energy balance between kinetic energies of large-scale flows and turbulence, thermal and magnetic energies in this model is also found to be in a good agreement with the flare energy balance derived from observations \citep[see e.g.][]{kone17}.

The twist visible in thermal EUV continuum and cross-loop polarisation gradient of microwave emission can be used for observational detection of twisted coronal loops. In fact, recently observations of CLPG have been reported by \citet{shae18}, although the effect appears to be weak due to low spatial resolution of the observations. On the other hand, correlation of plasma turbulence with plasma temperature as well as expansion of the HXR sources cannot be reliably used for the observational detection of twisted coronal loops, as these effects could also  be observed in other configurations. Still, the fact that our model produces these features indicates that it is quite realistic and can be further developed  for individual flare modelling. 

The MHDTP model thus has been very successful in modelling the acceleration and transport of energetic particles in solar flares, and a particular strength of the approach is that it can be used for a wide range of reconnection scenarios and can account for realistic global magnetic field configurations and dynamics. Obviously, the main drawback of the MHDTP method is the absence of feedback: particle energisation (or, in general, any deviation from Maxwellian distribution) does not affect electromagnetic field evolution determined by MHD. It can be ignored when the fraction  of energetic particles is low. In order to evaluate the validity of the method, particle acceleration efficiency can be used: the MHDTP method should provide reliable results if the amount of energy carried by accelerated particles is negligible compared to the amount of energy released during the whole reconnection event. 

\section{Hybrid MHD-kinetic models}

\subsection{Hybrid fluid-kinetic methods}

When the plasma distribution function deviates significantly from a Maxwellian, the effects of non-thermal plasma on the electro-magnetic field needs to be taken into account using self-consistent kinetic modelling. 
Particle-in-cell method is a powerful kinetic approach, which is widely applied to non-thermal plasma. In the last decade, this method has been widely used to study particle acceleration in the solar corona, heliosphere and magnetospheres. Notably \citet{drae06a,drae06} and \citet{drsw12}  showed that electrons and ions can be effectively accelerated by interacting with numerous contracting magnetic islands. Also, PIC has been used to study particle acceleration by propagating waves \citep[e.g.][]{gene04,tsik11}. A number of 2D and 3D models of magnetic reconnection in solar and space plasmas have been developed using PIC \cite[e.g.][]{hese01,hese02,prit08,tsha08}, see also \citet{bire12} for review. However, the explicit PIC method has severe limitations on the size of the simulation domain and timestep used in simulations imposed by the Debye scale and plasma frequency.

Without a  full kinetic treatment, non-thermal plasma can potentially be described using a hybrid approach. The term ``hybrid' refers to a broad  family of methods, in which  part of plasma  is described using a kinetic approach (Vlasov, PIC etc), while the  bulk  plasma is treated with a fluid model (usually MHD). A mathematical analysis of hybrid methods can be found in \citet{hole85} and \citet{morr98}; a comparison of a specific hybrid method with other methods for the forced magnetic reconnection model can be found in \citet{bire05}.

\begin{figure*}   
\centerline{\includegraphics[width=0.65\textwidth,clip=]{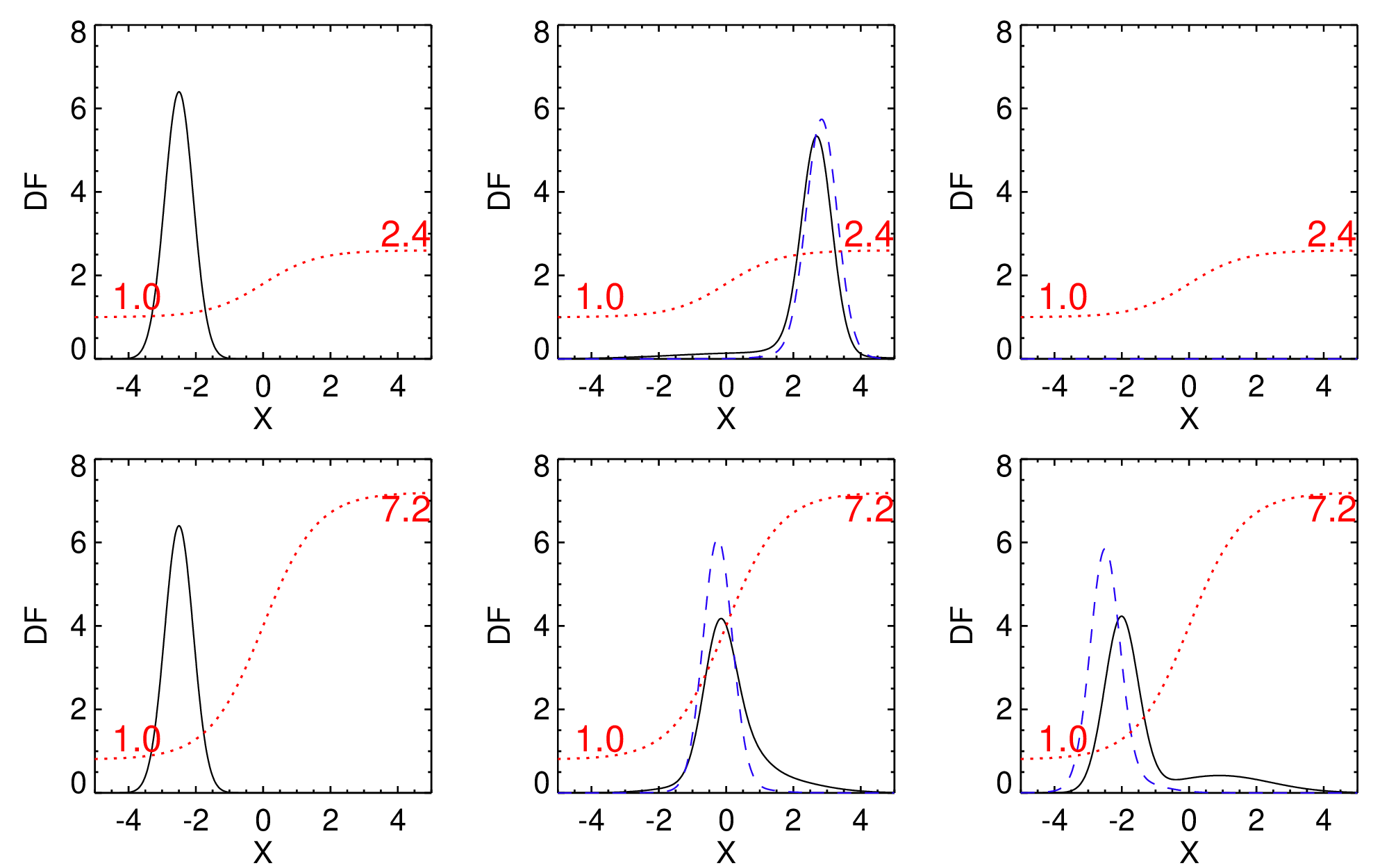}}
\caption{\scriptsize Comparison of distribution functions for  particles interacting with a magnetic mirror calculated using drift-kinetic (black solid lines) and reduced kinetics (blue dashed lines). Upper panels are for magnetic field contrast 2.4, lower panels are for field contrast 7.2. Left, middle and right panels show different instants in  time. \normalsize}
\end{figure*}

Before discussing fluid-kinetic methods, it is worth mentioning two-fluid and Hall magnetohydrodynamic approaches. Although they are completely ``fluid'' approaches, these methods treat electrons and ions separately, as well as use more parameters to describe the plasma ({\it e.g.} by using the pressure tensor, unlike in classical MHD, where pressure is described by a scalar function). As the result, they capture some elements of collisionless plasma physics beyond MHD, which are potentially relevant to solar flares. The Hall term in Ohm's Law becomes significant on scales comparable to the ion skin depth, about 10~m in the solar corona, which is comparable to current sheet widths predicted by resistive MHD, and hence should in principle be taken into account in reconnection simulations \citep{broe14}. Incorporating the Hall term in reconnection significantly increases the reconnection rate compared with single-fluid MHD, and thus the paralell electric field accelerating non-thermal particles \citep{broe14,mctu17}. MHDTP models incorporating two-fluid MHD should produce more realistic results, compared to single fluid MHD described in Section~2. Thus, recently a combination of two-fluid MHD and test-particle simulations has been used to explain the origin of non-thermal ions and electrons in the MAST spherical tokamak \citep{mctu17}

The most common hybrid models treat  ions as particles while  electrons are considered as charge-neutralising fluid. This approximation allows the study of  ion trajectories without resolving electron trajectories and electrostatic scales, making it possible to model domains with much larger spatial size (and also longer time-scales)  compared with  fully-kinetic models. This, in turn, provides the  opportunity to study  connections between phenomena at different spatial scales or, in other words,  kinetic effects behind MHD-scale phenomena. 

This kinetic-protons-fluid-electron approach has been extensively used for modelling reconnection and particle acceleration in the Earth magnetotail \citep[e.g.][]{nafu98,arsc01,kare04,omie09,aune11}. Although these  methods are often used in modelling of heliospheric and magnetospheric plasmas, there are not many such hybrid models of solar plasma. Indeed, this approach, while having some potential benefits, clearly cannot tackle the fundamental challenge of modelling solar flares with large populations of non-thermal electrons.  However, in recent years several hybrid models have been developed in context of solar flares. For example, \citet{gugi12} investigated particle acceleration on termination shocks formed by fast plasma outflow in reconnecting current sheets in the corona. Interestingly, in this study hybrid simulations are combined with test-particle calculations: while hybrid kinetic-proton-fluid-electron model produces proton trajectories, electron trajectories are calculated on the next step using full-trajectory test-particle method.

There are other, less commonly used hybrid methods. In  one approach, which is attracting increasing attention, kinetic and fluid methods are coupled geometrically, {\it i.e.} they describe separate parts of magnetic reconnection region. Both  Lagrangian and Eulerean kinetic methods are used. Thus, several recent  studies combine a  PIC model of the diffusion region with a two-fluid MHD model of peripherial regions of current sheet \citep[e.g.][]{suku07,dale14,make17}. In this case, on the boundary between kinetic and fluid regions, moments of distribution functions derived from PIC partcles are passed to the MHD region, while densities, temperatures and velocities are passed to the PIC region. Synchronising timesteps in MHD and kinetic simulations might be problematic, therefore, this methodology is most effective using an implicit (or semi-implicit)  PIC code which can use relatively large time-steps, since the time-step of the PIC and the MHD codes can then be matched \citep{make17}. A related approach is used by \citet{baue13}, who use the output of a restricted region of a 3D MHD simulation of a 3D reconnecting null to provide the initial and boundary  conditions for a PIC simulation. Similarly, MHD can be coupled with Vlasov simulations \citep[e.g.][]{riee15}. Using Vlasov can be more computationally expensive (especially in 3D cases); however, it makes it easier to stitch the  MHD and kinetic solutions. Generally, such coupled approaches can be extremely useful in some cases, such as magnetic reconnection in current sheets with regular structure and defined position, when the location of the strongly ``kinetic'' diffusion region is known. However, they  would not be effective for complex and fragmentary current structures (as discussed in Section 2 above), and cannot tackle the problem of modelling populations of non-thermal electrons which are observed in large volumes of the flaring corona.

\subsection{Reduced kinetic approach}

An approach being currently developed is the so-called ``reduced kinetic''  (RK) approach. This  combines a  drift-kinetic treatment of plasma transport along magnetic field with two-fluid MHD treatment of plasma transport across magnetic field. The idea of this approach stems from previous MHDTP models showing that in the absence of strong magnetic mirroring in the system, energetic particles are collimated along magnetic field lines \citep[e.g.][]{gore14}. This is motivated by the fact that parallel electric field is the main acceleration factor in most MHDTP models (see Section 2 above). If particles are adiabatic then, in the absence of strong magnetic field curvature, this electric field component affects only parallel particle velocities, while the  perpendicular velocity ({\it i.e.} Larmor gyro-velocity) distribution remains nearly thermal. The main benefit of this approximation is to reduce the dimensionality of the phase space. Thus, there are two velocity dimensions in normal drift-kinetic approach, for instance $[\vec{r}, v_{||}, v_g]$ (although, gyro-velocity can be substituted by magnetic moment, pitch-angle etc), while in the reduced-kinetic approach,  particles may have arbitrary distribution only for parallel velocity, {\it i.e.} $[\vec{r}, v_{||}]$.

In the RK approach, plasma is described by distribution functions for each species $f(\vec{r}, v_{||})$, while perpendicular velocity distribution is characterised by function $\tau(\vec{r}, v_{||})$, its physical meaning is close to the perpendicular temperature. This is equivalent to ``fixing'' a part of the distribution function as follows:

\[
F=\frac {d^3 n}{d\vec{r} dv_{||} d v^2_g}= f(\vec{r}, v_{||}) \exp \left( -\frac{v_g^2}{\tau(\vec{r}, v_{||})}\right).
\]
Substituting this into drift-kinetic equations yields two equations:
\begin{eqnarray}
\frac {\partial F_s}{\partial t} &=&- (\vec{V}_{s}+ v_{||} \vec{b}) \frac {\partial F_s}{\partial \vec{r}}  - F_s \vec{\nabla} \cdot \vec{V} \nonumber \\
&& - \frac qm \vec{\mathcal{E}}\cdot \vec{b} \frac {\partial F_s}{\partial v_{||}} +  G \tau \frac {\partial F_s}{\partial v_{||}} +\mathcal{S}_s \nonumber \\
\frac {\partial \tau_s}{\partial t} &=& - (\vec{V}_{s}+ v_{||} \vec{b}) \frac {\partial \tau_s}{\partial \vec{r}} - \tau_s \vec{\nabla} \cdot \vec{V} \nonumber \\
&& - \frac qm \vec{\mathcal{E}}\cdot \vec{b} \frac {\partial \tau_s}{\partial v_{||}} +  G \tau \frac {\partial \tau_s}{\partial v_{||}} + 2 G v_{||} \tau_s+\mathcal{T}_s, \nonumber
\end{eqnarray}
where $G = \frac {\vec{\nabla} B \cdot \vec{b}}{B}$, $\vec{\mathcal{E}}$ is electric field, $\vec{V}_{s}$ is bulk flow velocity. $\mathcal{S}_s$ and $\mathcal{T}_s$ are the terms responsible for particle-particle and wave-particle interaction. These terms are very important for modelling non-ideal effects, such as magnetic reconnection and current dissipation. In principle, they have to be derived using the collisional operators (e.g. diffusion tensor and the friction force terms) in the drift-kinetic equations. Alternatively, a simplified parametric form, mimicking the effect of resistivity in the induction equation in MHD, can be used. 

It is important to note, that the parameters used in these terms, generally, are not calculated self-consistently, unlike in kinetic models, resolving small, ``kinetic'' scales. Therefore, this large-scale hybrid approach has an important drawback: similar to the resistivity, required in non-ideal MHD simulations, RK approach requires parameters accounting for collisions and wave-particle interaction.

If the parallel velocity distribution is Maxwellian, these equations reduce to the mass conservation and energy equations used in MHD. The main source of error for this approach is strong magnetic field variation along particle paths, in other words high values of $(\vec{b} \cdot \vec{\nabla}) B$, since this transforms parallel into perpendicualr velocities. In order to establish the range of $(\vec{b} \cdot \vec{\nabla}) B$ values where this method is still viable, we test the above equations using 1D magnetic mirror model. The results are shown in Figure 2. Essentially, it shows that the reduced kinetic method can provide  good results when the system does not have strong magnetic mirrors.

The main goal of using reduced kinetics is to speed up computational modelling of partially non-thermal plasma at large spatial and temporal scales. This, in turn, means that electrostatic scales are not resolved and, therefore, the ``quasi-neutral'', lower-order form Maxwell equations need to be used \citep{trca15}. This  can be derived assuming that net charge density and the displacement current are zero everywhere:
\begin{eqnarray}
\vec{\nabla} \times \vec{E} = - \frac{\partial \vec{B}}{\partial t} \nonumber \\
\vec{\nabla} \times \vec{B} = \mu \vec{j}, \nonumber
\end{eqnarray}
where current density is calculated as a moment of the distribution function. Although this approach is limited to configurations with rather smooth magnetic fields, where magnetic mirroring is weak, it can, potentially, have a number of applications from 2D fluid-kinetic models of reconnecting current sheets to cylindrically-symmetric models of charged particle transport in the solar corona.

\section{Summary}

Full understanding of solar flares requires a the development of models which can encompass the vast range of relevant spatial and temporal scales, from large global scales which are well described by fluid models, to the small plasma scale-lengths involving kinetic physics. 

The MHDTP approach is a simple, robust tool that can be used for modelling non-thermal plasma in the corona, as well as in the heliosphere.  It is particularly useful for studying particle acceleration and transport in large-scale and complex magnetic configurations, and also for developing basic understanding of the processes of particle acceleration and their dependences on the electromagnetic field. That is why it is widely used not only to develop generic flare models, but also for modelling individual flaring events, the response of the heliosphere and space weather prediction \citep[see e.g.][]{lape13}.

The lack of self-consistency in the main drawback of the MHDTP approach. Hence, plasmas with a high number of non-thermal particles  -  as in  stronger solar flares - require either a full kinetic or hybrid fluid-kinetic treatment. Although hybrid methods have fewer restrictions compared with  full kinetic methods, they are still computationally expensive and currently are not commonly used for large-scale flare modelling with realistic parameters. However, fast progress in available computational resources offered by new high-performance computing facilities, as well as the development of new techniques,  can make hybrid approach an essential tool in complex solar flare modelling.

It should be noted that kinetic phenomena can be studied at increasingly larger scales even without hybrid methods, and advances in computer power and numerical techniques are extending the reach of kinetic models \cite[e.g.][]{daue11,lape17}. Thus, implicit integration schemes in PIC simulations make it possible to avoid resolving electrostatic scales \citep{lape17,puce17}. Furthermore, the gyrokinetic approach, widely-used for modelling fusion plasmas, offers potential for modelling magnetic reconnection in strongly-magnetised plasmas, which is generally relevant in solar flares \citep{tene14}.

Finally, it needs to be noted that, although many coronal models of magnetic reconnection ignore electrostatic scales, these scales may be important, particularly in the outer corona and heliosphere. Plasma oscillations induced by propagating electron beams are responsible for producing low-frequency radio-emission, which is another essential tool for observational diagnostics of flares and related events \citep[see e.g.][]{base98}. Furthermore, interaction between plasma waves and non-thermal electrons affect both electron and radio-wave propagation, thus, affecting spatial and temporal properties of low-frequency radio emission sources \citep{kone17n}, and, hence, plasma wave effects may need to be incorporated into large-scale fluid-kinetic solar flare models.

{\it \bf Acknowledgement.} MG and PKB are grateful to STFC for support from Consolidated Grant ST/000428/1


\begin{thebibliography}{}
\bibitem[Alexander et al.(2006)]{alee06} Alexander, D., Liu, R., Gilbert, H.R., 2006. Hard X-Ray Production in a Failed Filament Eruption, Astrophys. J., 653, 719-724. http://doi.org/10.1086/508137
\bibitem[Arber et al.(2001)]{arbe01} Arber, T.D., Longbottom, A.W., Gerrard, C.L., Milne, A.M., 2001. A Staggered Grid, Lagrangian-Eulerian Remap Code for 3-D MHD Simulations, J.Comp.Phys., 171, 151-181. http://doi.org/10.1006/jcph.2001.6780
\bibitem[Arzner \& Scholer(2001)]{arsc01} Arzner, K., Scholer, M., 2001. Kinetic structure of the post plasmoid plasma sheet during magnetotail reconnection, J. Geophys. Res., 106, 3827-3844. http://doi.org/10.1029/2000JA000179
\bibitem[Arzner and Vlahos(2004)]{arvl04} Arzner,K., Vlahos,L., 2004. Particle Acceleration in Multiple Dissipation Regions, Astrophys. J. Lett, 605, L69-L72. http://doi.org/10.1086/392506
\bibitem[Aschwanden(2002)]{asch02} Aschwanden, M.J. 2002. Particle acceleration and kinematics in solar flares - A Synthesis of Recent Observations and Theoretical Concepts (Invited Review), Space Sci. Rev, 1-227. http://doi.org/10.1023/A:1019712124366
\bibitem[Aschwanden et al.(2009)]{asce09} Aschwanden, M.J., Wuelser, J.P., Nitta, N.V., Lemen, J.R., 2009. Solar Flare and CME Observations with STEREO/EUV, Solar Phys., 256, 3-40. http://doi.org/10.1007/s11207-009-9347-4
\bibitem[Aunai et al.(2011)]{aune11} Aunai, N., Belmont, G., Smets, R., 2011. Proton acceleration in antiparallel collisionless magnetic reconnection: Kinetic mechanisms behind the fluid dynamics, J. Geophys. Res., 116, ID A09232. http://doi.org/10.1029/2011JA016688
\bibitem[Bareford et al.(2016)]{bare16} Bareford,M.R., Gordovskyy,M., Browning,P.K., Hood,A.W., 2016. Energy Release in Driven Twisted Coronal Loops, Solar Phys., 291, 187-209. http://doi.org/10.1007/s11207-015-0824-7
\bibitem[B\'{a}rta et al.(2010)]{bare10} B\'{a}rta, M., B\"{u}chner, J., Karlick\'{y}, M., 2010. Multi-scale MHD approach to the current sheet filamentation in solar coronal reconnection, Adv. Sp. Res., 45, 10-17. http://doi.org/10.1016/j.asr.2009.07.025
\bibitem[B\'{a}rta et al.(2011)]{bare11} B\'{a}rta, M., B\"{u}chner, J., Karlick\'{y}, M., Sk\'{a}la, J., 2011. Spontaneous Current-layer Fragmentation and Cascading Reconnection in Solar Flares. I. Model and Analysis, Astrophys. J., 737, ID 24. http://doi.org/10.1088/0004-637X/737/1/24
\bibitem[Bastian et al.(1998)]{base98} Bastian, T.S., Benz, A.O., Gary, D.E., 1998. Radio Emission from Solar Flares, Ann.Rev.Astron. Astrophys., 36, 131-188. http://doi.org/10.1146/annurev.astro.36.1.131
\bibitem[Baumann et al.(2013)]{baue13} Baumann, G., Haugb{\o}lle, T., Nordlund, {\AA}., 2013. Kinetic Modeling of Particle Acceleration in a Solar Null-point Reconnection Region, Astrophys. J., 771, ID 93. http://doi.org/10.1088/0004-637X/771/2/93
\bibitem[Benz(2017)]{benz17} Benz, A.O., 2017. Flare Observations, Liv.Rev.Solar Phys., 14, ID 2. http://doi.org/10.1007/s41116-016-0004-3
\bibitem[Birn et al.(2004)]{bire04} Birn, J., Thomsen, M.F., Hesse, M., 2004. Electron acceleration in the dynamic magnetotail: Test particle orbits in three-dimensional magnetohydrodynamic simulation fields, Phys.Plasm., 11, 1825-1833. http://doi.org/10.1063/1.1704641
\bibitem[Birn et al.(2005)]{bire05} Birn, J., Galsgaard, K., Hesse, M., Hoshino, M., Huba, J., Lapenta, G., Pritchett, P.L., Schindler, K., Yin, L., Büchner, J., Neukirch, T., Priest, E.R., 2005. Forced magnetic reconnection, Geophys.Res.Lett., 32, ID L06105. http://doi.org/10.1029/2004GL022058
\bibitem[Birn et al.(2012)]{bire12} Birn, J., Artemyev, A.V., Baker, D.N., Echim, M., Hoshino, M., Zelenyi, L.M., 2012. Particle Acceleration in the Magnetotail and Aurora, Space Sci. Rev., 173, 49-102. http://doi.org/10.1007/s11214-012-9874-4
\bibitem[Birn et al.(2017)]{bire17} Birn, J., Battaglia, M., Fletcher, L., Hesse, M. \& Neukirch,T., 2017. Can Substorm Particle Acceleration Be Applied to Solar Flares?, Astrophys. J., 848, ID 116. http://doi.org/10.3847/1538-4357/aa8ad4
\bibitem[Botha et al.(2012)]{bote12} Botha, G.J.J., Arber, T.D., Srivastava, A.K., 2012. Observational Signatures of the Coronal Kink Instability with Thermal Conduction, Astrophys. J., 745, ID 53. http://doi.org/10.1088/0004-637X/745/1/53
\bibitem[Borissov et al.(2017)]{bore17} Borissov,A., Kontar, E.P., Threlfall, J. \& Neukirch, T. 2017. Particle acceleration with anomalous pitch angle scattering in 2D magnetohydrodynamic reconnection simulations, Astron. Astrophys., 605, ID A73. http://doi.org/10.1051/0004-6361/201731183
\bibitem[Brown et al.(2009)]{broe09} Brown, J.C., Turkmani, R., Kontar, E.P., MacKinnon, A.L., Vlahos, L., 2009. Local re-acceleration and a modified thick target model of solar flare electrons, Astron. Astrophys., 508, 993-1000. http://doi.org/10.1051/0004-6361/200913145
\bibitem[Browning and Vekstein(2001)]{brve01} Browning,P.K., Vekstein,G.E., 2001. Particle acceleration at an X-type reconnection site with a parallel magnetic field, J. Geophys. Res. Space Physics, 106, 18677-18692. http://doi.org/10.1029/2001JA900014
\bibitem[Browning \& Van der Linden(2003)]{brva03} Browning, P.K., Van der Linden, R.A.M., 2003. Solar coronal heating by relaxation events, Astron. Astrophys., 400, 355-367. http://doi.org/10.1051/0004-6361:20021887
\bibitem[Browning et al.(2008)]{broe08} Browning, P.K., Gerrard, C., Hood, A.W., Kevis, R., van der Linden, R.A.M., 2008. Heating the corona by nanoflares: simulations of energy release triggered by a kink instability, Astron. Astrophys., 485, 837-848. http://doi.org/10.1051/0004-6361:20079192
\bibitem[Browning et al.(2010)]{broe10} Browning, P.K., Dalla, S., Peters, D., Smith, J., 2010. Scaling of particle acceleration in 3D reconnection at null points, Astron. Astrophys., 520, ID 105. http://doi.org/10.1051/0004-6361/201014964
\bibitem[Browning et al.(2014)]{broe14} Browning, P.K., Stanier, A., Ashworth, G., McClements, K.G., Lukin, V.S., 2014. Self-organization during spherical torus formation by flux rope merging in the Mega Ampere Spherical Tokamak, Plas.Phys.Conf.Fus., 56, ID 064049. http://doi.org/10.1088/0741-3335/56/6/064009
\bibitem[Burge et al.(2014)]{bure14} Burge,C.A., MacKinnon, A.L. \& Petkaki,P., 2014. Effect of binary collisions on electron acceleration in magnetic reconnection, Astron. Astrophys., 561, ID 107. http://doi.org/10.1051/0004-6361/201322199
\bibitem[Cargill et al.(2006)]{care06} Cargill, P.S., Vlahos, L., Turkmani, R., Galsgaard, K., Isliker, H., 2006. Particle acceleration in a three-dimensional model of reconnecting coronal magnetic fields, Space Sci. Rev., 124, 249-259. http://doi.org/10.1007/s11214-006-9108-8
\bibitem[Cargill et al.(2012)]{care12} Cargill, P.S., Vlahos, L., Baumann, G., Drake, J.F., Nordlund, {\AA}., 2012. Current Fragmentation and Particle Acceleration in Solar Flares, Space Sci. Rev, 173, 223-245. http://doi.org/10.1007/s11214-012-9888-y
\bibitem[Chen(2011)]{chen11} Chen, P.F., 2011. Coronal Mass Ejections: Models and Their Observational Basis, Liv.Rev.Solar Phys., 8, 1. http://doi.org/10.12942/lrsp-2011-1
\bibitem[Daldorff et al.(2014)]{dale14} Daldorff, L.K.S., T\'{o}th, G., Gombosi, T.I, Lapenta, G., Amaya, J., Markidis, S., Brackbil, J.U., 2014. Two-way coupling of a global Hall magnetohydrodynamics model with a local implicit particle-in-cell model, J. Comp. Phys., 268, 236-254. http://doi.org/10.1016/j.jcp.2014.03.009
\bibitem[Dalla \& Browning(2005)]{dabr05} Dalla, S., Browning, P.K., 2005. Particle acceleration at a three-dimensional reconnection site in the solar corona, Astron. Astrophys., 436, 1103-1111. http://doi.org/10.1051/0004-6361:20042589
\bibitem[Dalla \& Browning(2006)]{dabr06} Dalla, S., Browning, P.K., 2006. Jets of Energetic Particles Generated by Magnetic Reconnection at a Three-dimensional Magnetic Null, Astrophys. J. Lett., 640, L99-L102. http://doi.org/10.1086/503302
\bibitem[Dalla \& Browning(2008)]{dabr08} Dalla, S., Browning, P.K., 2008. Particle trajectories and acceleration during 3D fan reconnection, Astron. Astrophys., 491, 289-295. http://doi.org/10.1051/0004-6361:200809771
\bibitem[Daughton et al (2011)]{daue11} Daughton, W., Roytershteyn, V.,  Karimabadi, H.,  Yin, L., Albright, B.J., Bergen, B., Bowers,K.J., 2011. Role of electron physics in the development of turbulent magnetic reconnection in collisionless plasmas, Nature Phys., 7, 539-542. http://doi.org/10.1038/nphys1965
\bibitem[Dmitruk et al.(2004)]{dmie04} Dmitruk,P., Matthaeus,W.H., Seenu,N., 2004. Test Particle Energization by Current Sheets and Nonuniform Fields in Magnetohydrodynamic Turbulence, Astrophys. J., 617, 677-679. http://doi.org/10.1086/425301
\bibitem[Drake \& Swisdak(2012)]{drsw12} Drake, J.F., Swisdak, M., 2012. Ion heating and acceleration during magnetic reconnection relevant to the corona, Space Sci. Rev., 172, 227-240. http://doi.org/10.1007/s11214-012-9903-3
\bibitem[Drake et al.(2006a)]{drae06a} Drake, J.F., Swisdak, M., Schoeffler, K.M., Rogers, B.N., Kobayashi, S., 2006. Formation of secondary islands during magnetic reconnection, Geophys. Res. Lett., 33, ID L13105. http://doi.org/10.1029/2006GL025957
\bibitem[Drake et al.(2006b)]{drae06} Drake, J.F., Swisdak, M., Che, H., and Shay, M.A., 2006. Electron acceleration from contracting magnetic islands during reconnection, Nature, 443, 553-556. http://doi.org/10.1038/nature05116
\bibitem[Doschek et al.(2007)]{dose07} Doschek, G.A., Mariska, J.T., Warren, H.P., Brown, C.M., Culhane, J.L., Hara, H., Watanabe, T., Young, P.R., Mason, H.E., 2007. Nonthermal Velocities in Solar Active Regions Observed with the Extreme-Ultraviolet Imaging Spectrometer on Hinode, Astrophys. J. Lett., 667, L109-L112. http://doi.org/10.1086/522087
\bibitem[Doschek et al.(2008)]{dose08} Doschek, G.A., Warren, H.P., Mariska, J.T., Muglach, K., Culhane, J.,L., Hara, H., Watanabe, T., 2008. Flows and Nonthermal Velocities in Solar Active Regions Observed with the EUV Imaging Spectrometer on Hinode: A Tracer of Active Region Sources of Heliospheric Magnetic Fields?, Astrophys. J., 686, 1362-1371. http://doi.org/10.1086/591724
\bibitem[Fleishman \& Kuznetsov(2010)]{flku10} Fleishman, G.D., Kuznetsov, A.A., 2010. Fast Gyrosynchrotron Codes, Astrophys. J., 721, 1127-1141 http://doi.org/10.1088/0004-637X/721/2/1127
\bibitem[G\'{e}not et al.(2004)]{gene04} G\'{e}not, V., Louarn, P., Mottez, F., 2004. Alfv\'{e}n wave interaction with inhomogeneous plasmas: acceleration and energy cascade towards small-scales, Ann. Geophys., 22, 2080-2096. http://doi.org/10.5194/angeo-22-2081-2004
\bibitem[Gibson et al.(2006)]{gibe06} Gibson, S.E., Fan, Y., T{\"o}r{\"o}k, T., Kliem, B., 2006. The Evolving Sigmoid: Evidence for Magnetic Flux Ropes in the Corona Before, During, and After CMES, Space Sci. Rev, 124, 131-144. http://doi.org/10.1007/s11214-006-9101-2
\bibitem[Giuliani et al.(2005)]{giue05} Giuliani, P., Neukirch, T., Wood, P., 2005. Particle Motion in Collapsing Magnetic Traps in Solar Flares. I. Kinematic Theory of Collapsing Magnetic Traps, Astrophys. J., 635, 636-646. http://doi.org/10.1086/497366
\bibitem[Gordovskyy et al.(2010a)]{gore10a} Gordovskyy, M., Browning, P.K., Vekstein, G.E., 2010a. Particle acceleration in a transient magnetic reconnection event, , Astron. Astrophys., 519, ID A21. http://doi.org/10.1051/0004-6361/200913569
\bibitem[Gordovskyy et al.(2010b)]{gore10b} Gordovskyy, M., Browning, P.K., Vekstein, G.E., 2010b. Particle Acceleration in Fragmenting Periodic Reconnecting Current Sheets in Solar Flares, , Astrophys.J., 720, 1603-1611. http://doi.org/10.1088/0004-637X/720/2/1603
\bibitem[Gordovskyy \& Browning(2011)]{gobr11} Gordovskyy, M., Browning, P.K., 2011. Particle Acceleration by Magnetic Reconnection in a Twisted Coronal Loop, Astrophys. J., 729, ID 101. http://doi.org/10.1088/0004-637X/729/2/101
\bibitem[Gordovskyy \& Browning(2012)]{gobr12} Gordovskyy, M., Browning, P.K., 2012. Magnetic Relaxation and Particle Acceleration in a Flaring Twisted Coronal Loop, Solar Phys., 277, 299-316. http://doi.org/10.1007/s11207-011-9900-9
\bibitem[Gordovskyy et al.(2013)]{gore13} Gordovskyy, M., Browning, P.K., Kontar, E.P., Bian, N.H., 2013. Effect of Collisions and Magnetic Convergence on Electron Acceleration and Transport in Reconnecting Twisted Solar Flare Loops, Solar Phys., 284, 489-498. http://doi.org/10.1007/s11207-012-0124-4
\bibitem[Gordovskyy et al.(2014)]{gore14} Gordovskyy, M., Browning, P.K., Kontar, E.P., Bian, N.H., 2014. Particle acceleration and transport in reconnecting twisted loops in a stratified atmosphere, Astron. Astrophys., 561, ID A72. http://doi.org/10.1051/0004-6361/201321715
\bibitem[Gordovskyy et al.(2016)]{gore16} Gordovskyy, M., Kontar, E.P., Browning, P.K., 2016. Plasma motions and non-thermal line broadening in flaring twisted coronal loops, Astron. Astrophys., 589, ID A104. http://doi.org/10.1051/0004-6361/201527249
\bibitem[Gordovskyy et al.(2017)]{gore17} Gordovskyy, M., Browning, P.K., Kontar, E. P., 2017. Polarisation of microwave emission from reconnecting twisted coronal loops, Astron. Astrophys., 604, ID A116. http://doi.org/10.1051/0004-6361/201629334
\bibitem[Grady \& Neukirk(2009)]{grne09} Grady, K.J., Neukirch, T., 2009. An extension of the theory of kinematic MHD models of collapsing magnetic traps to 2.5D with shear flow and to 3D, Astron. Astrophys., 508, 1461-1468. http://doi.org/10.1051/0004-6361/200913230
\bibitem[Grady et al.(2012)]{grae12} Grady, K.J., Neukirch, T., Giuliani, P., 2012. A systematic examination of particle motion in a collapsing magnetic trap model for solar flares, Astron. Astrophys., 546, ID A85. http://doi.org/10.1051/0004-6361/201218914
\bibitem[Graf von der Pahlen \& Tsiklauri(2016)]{grts16} Graf von der Pahlen, J., Tsiklauri, D., 2016. Role of electron inertia and reconnection dynamics in a stressed X-point collapse with a guide-field, Astron. Astrophys., 595, ID A84. http://doi.org/10.1051/0004-6361/201628071
\bibitem[Guo \& Giacalone(2012)]{gugi12} Guo, F., Giacalone, J., 2012. Particle Acceleration at a Flare Termination Shock: Effect of Large-scale Magnetic Turbulence, Astrophys. J., 753, ID 28. http://doi.org/10.1088/0004-637X/753/1/28
\bibitem[Guo et al.(2010)]{guoe10} Guo, J.-N., B{\"u}chner, J., Otto, A., Santos, J., Marsch, E., Gan, W.-Q., 2010. Is the 3-D magnetic null point with a convective electric field an efficient particle accelerator?, Astron. Astrophys., 513, ID A73. http://doi.org/10.1051/0004-6361/200913321
\bibitem[Guo et al.(2014)]{guoe14} Guo, X., Sironi, L., and Narayan, R., 2014. Non-thermal Electron Acceleration in Low Mach Number Collisionless Shocks. I. Particle Energy Spectra and Acceleration Mechanism, Astrophys. J., 794, ID 153. http://doi.org/10.1088/0004-637X/794/2/153
\bibitem[Hahm \& Kulsrud(1985)]{haku85} Hahm, T.S., Kulsrud, R.M., 1985. Forced magnetic reconnection, Phys. Plasmas, 28, 2412-2418. http://doi.org/10.1063/1.865247
\bibitem[Hamilton et al.(2005)]{hame05} Hamilton, B., Fletcher, L., McClements, K.G., Thyagaraja, A., 2005. Electron Acceleration at Reconnecting X-Points in Solar Flares, Astrophys. J., 625, 496-505. http://doi.org/10.1086/430100
\bibitem[Hannah \& Fletcher(2006)]{hafl06} Hannah, I.G., Fletcher, L., 2006. Comparison of the Energy Spectra and Number Fluxes From a simple Flare Model to Observations, Solar Phys., 236, 59-74. http://doi.org/10.1007/s11207-006-0139-9
\bibitem[Hesse et al.(2001)]{hese01} Hesse, M., Kuznetsova, M., Birn, J., 2001. Particle-in-cell simulations of three-dimensional collisionless magnetic reconnection, J. Geophys. Res., 106,  29831-29842. http://doi.org/10.1029/2001JA000075
\bibitem[Hesse et al.(2002)]{hese02} Hesse, M., Kuznetsova, M., Hoshino, M., 2002, The structure of the dissipation region for component reconnection: Particle simulations 29, ID 1563. http://doi.org/10.1029/2001GL014714
\bibitem[Hewett et al.(1988)]{hewe88} Hewett, D.W., Frances, G.E., Max, C.E., 1988. New regimes of magnetic reconnection in collisionless plasmas, Phys.Rev.Lett, 61, 893-896. http://doi.org/10.1103/PhysRevLett.61.893
\bibitem[Holm et al.(1985)]{hole85} Holm, D.D., Marsden, J.E., Ratiu, T.S., Weinstein, A., 1985. Nonlinear stability of fluid and plasma equilibria, Phys.Rep., 123, 1-116. http://doi.org/10.1016/0370-1573(85)90028-6
\bibitem[Hood et al.(2009)]{hooe09} Hood, A.W., Browning, P.K., van der Linden, R.A.M., 2009. Coronal heating by magnetic reconnection in loops with zero net current, Astron. Astrophys., 506, 913-925. http://doi.org/10.1051/0004-6361/200912285
\bibitem[Hood et al.(2016)]{hooe16}  Hood, A.W., Cargill,P.C.,  Browning, P.K., Tam, K.V., 2016. An MHD Avalanche in a Multi-threaded Coronal Loop, Astrophys. J., 817, ID 5. http://doi.org/10.3847/0004-637X/817/1/5
\bibitem[Hudson \& Ryan(1995)]{hury95} Hudson, H., Ryan, J., 1995. High-Energy Particles In Solar Flares, Ann.Rev.Astron. Astrophys., 22, 239-282. http://doi.org/10.1146/annurev.aa.33.090195.001323
\bibitem[Isliker et al.(2017)]{isle17} Isliker, H., Pisokas, T., Vlahos, L., Anastasiadis, A., 2017. Particle Acceleration and Fractional Transport in Turbulent Reconnection, Astrophys. J., 849, ID 35. http://doi.org/10.3847/1538-4357/aa8ee8
\bibitem[Karimabadi et al.(2004)]{kare04} Karimabadi, H., Huba, J.D., Krauss-Varban, D., Omidi, N., 2004. On the generation and structure of the quadrupole magnetic field in the reconnection process: Comparative simulation study, Geophys.Res.Lett., 31, ID L07806. http://doi.org/10.1029/2004GL019553
\bibitem[Karlicky \& Barta(2006)]{kaba06} Karlick{\'y}, M., B{\'a}rta, M., 2006. X-Ray Loop-Top Source Generated by Processes in a Flare Collapsing Trap, Astrophys. J., 647, 1472-1479. http://doi.org/10.1086/505460
\bibitem[Klein \& Dalla(2017)]{klda17} Klein, K.-L., Dalla, S., 2017. Acceleration and Propagation of Solar Energetic Particles, Space Sci. Rev, 212, 1107-1136. http://doi.org/10.1007/s11214-017-0382-4
\bibitem[Kliem(1994)]{klie94} Kliem, B., 1994. Particle orbits, trapping, and acceleration in a filamentary current sheet model, Astrophys. J. Suppl., 90, 719-728. http://doi.org/10.1086/191896
\bibitem[Kontar et al.(2011)]{kone11} Kontar, E.P., Hannah, I.G., Bian, N.H., 2011. Acceleration, Magnetic Fluctuations, and Cross-field Transport of Energetic Electrons in a Solar Flare Loop, Astrophys. J., 730, ID L22. http://doi.org/10.1088/2041-8205/730/2/L22
\bibitem[Kontar et al.(2014)]{kone14} Kontar, E.P., Bian, N.H., Emslie, A.G., Vilmer, N., 2014. Turbulent Pitch-angle Scattering and Diffusive Transport of Hard X-Ray-producing Electrons in Flaring Coronal Loops, Astrophys. J., 780, ID 176. http://doi.org/10.1088/0004-637X/780/2/176
\bibitem[Kontar et al.(2017a)]{kone17} Kontar, E.P., Perez, J.E., Harra, L.K., Kuznetsov, A.A., Emslie, A.G., Jeffrey, N.L.S., Bian, N.H., Dennis, B.R., 2017. Turbulent Kinetic Energy in the Energy Balance of a Solar Flare, Phys.Rev.Lett., 118, ID 155101. http://doi.org/10.1103/PhysRevLett.118.155101
\bibitem[Kontar et al.(2017b)]{kone17n} Kontar, E.P., Yu, S., Kuznetsov, A.A., Emslie, A.G., Alcock, B., Jeffrey, N.L.S., Melnik, V.N., Bian, N.H., Subramanian, P., 2017. Imaging spectroscopy of solar radio burst fine structures, Nat.Comm., 8, ID 1515. http://doi.org/10.1038/s41467-017-01307-8
\bibitem[Kuznetsov et al.(2015)]{kuze15} Kuznetsov, A.A., Van Doorsselaere, T., Reznikova, V.E., 2015. Simulations of Gyrosynchrotron Microwave Emission from an Oscillating 3D Magnetic Loop, Solar Phys., 290, 1173-1194. http://doi.org/10.1007/s11207-015-0662-7
\bibitem[Lapenta et al.(2017)]{lape17} Lapenta, G., Gonzalez-Herrero, D., Boella, E., 2017. Multiple-scale kinetic simulations with the energy conserving semi-implicit particle in cell method, J.Plasm.Phys., 23, ID 705830205. http://doi.org/10.1017/S0022377817000137
\bibitem[Lapenta et al.(2013)]{lape13} Lapenta, G., Pierrard, V., Keppens, R., Markidis, S., Poedts, S., Sebek, O., Travnicek, P.M., Henri, P., Califano, F., Pegoraro, F., Faganello, M., Olshevsky, V., Restante, A.L., Nordlund, {\AA}., Trier, F.J., Mackay, D.H., Parnell, C.E., Bemporad, A., Susino, R., Borremans, K., 2013. SWIFF: Space weather integrated forecasting framework, J.Sp.Weath.Sp.Clim., 3, ID A05. http://doi.org/10.1051/swsc/2013027
\bibitem[Li et al.(2014)]{lie14} Li, T.C., Drake, J.F., Swisdak, M., 2014. Dynamics of Double Layers, Ion Acceleration, and Heat Flux Suppression during Solar Flares, Astrophys. J., 793, ID 7. http://doi.org/10.1088/0004-637X/793/1/7
\bibitem[Li et al.(2017)]{lie17} Li, Y.,  Wu, N., Lin, J., 2017. Charged-particle acceleration in a reconnecting current sheet including multiple magnetic islands and a nonuniform background magnetic field, Astron. Astrophys., 605, ID A120. http://doi.org/10.1051/0004-6361/201630026
\bibitem[Litvinenko(1996)]{litv96} Litvinenko, Yu.E,  1996. Particle orbits, trapping, and acceleration in a filamentary current sheet model, Astrophys.J., 462, 997-1004. http://doi.org/10.1086/177213
\bibitem[Litvinenko and Somov(1993)]{liso93} Litvinenko, Yu.E, Somov, B.V.,  1993. Particle acceleration in reconnecting current sheets, Solar Phys., 146, 127-133. http://doi.org/10.1007/BF00662174
\bibitem[Makwana et al.(2017)]{make17} Makwana, K.D, Keppens, R., Lapenta, G., 2017. Two-way coupling of magnetohydrodynamic simulations with embedded particle-in-cell simulations, Comp. Phys. Comm., 221, 81-94. http://doi.org/10.1016/j.cpc.2017.08.003
\bibitem[Masson et al.(2009)]{mase09} Masson, S., Pariat, E., Aulaniar, G., Schrijver,C.J., 2009. The Nature of Flare Ribbons in Coronal Null-Point Topology, Astrophys. J., 700, 559-578. http://doi.org/10.1088/0004-637X/700/1/559
\bibitem[McClements et al.(1990)]{mcce90} McClements, K.G., Su, J.J., Bingham, R., Dawson, J.M., Spicer, D.S., 1990. Simulation studies of electron acceleration by ion ring distributions in solar flares, Solar Phys., 130, 229-241. http://doi.org/10.1007/BF00156791
\bibitem[McClements and Turnyanskiy(2017)]{mctu17} McClements, K.G., Turnyanskiy, M.R., 2017. Energetic particles in laboratory, space and astrophysical plasmas, Plas. Phys. Cont. Fus., 59, ID 014012. http://doi.org/10.1088/0741-3335/59/1/014012
\bibitem[Morrison(1998)]{morr98} Morrison, P.J., 1998. Hamiltonian description of the ideal fluid, Rev.Mod.Phys., 70, 467-521. http://doi.org/10.1103/RevModPhys.70.467
\bibitem[Nakamura \& Fujimoto(1998)]{nafu98} Nakamura, M.S., Fujimoto, M., 1998. A three-dimensional hybrid simulation of magnetic reconnection, Geophys.Res.Lett., 25, 2917-2920. http://doi.org/10.1029/98GL02154
\bibitem[Nakariakov \& Melnikov(2009)]{name09} Nakariakov, V.M., Melnikov, V.F., 2009. Quasi-Periodic Pulsations in Solar Flares, Space Sci. Rev, 149, 119-151. http://doi.org/10.1007/s11214-009-9536-3
\bibitem[Nita et al.(2015)]{nite15} Nita, G.M., Fleishman, G.D., Kuznetsov, A.A., Kontar, E.P., Gary, D.E., 2015. Three-dimensional Radio and X-Ray Modeling and Data Analysis Software: Revealing Flare Complexity, Astrophys. J., 799, ID 236. http://doi.org/10.1088/0004-637X/799/2/236	
\bibitem[Omidi et al.(2009)]{omie09} Omidi, N., Phan, T., Sibeck, D.G., 2009. Hybrid simulations of magnetic reconnection initiated in the magnetosheath, J. Geophys. Res., 114, ID A02222. http://doi.org/10.1029/2008JA013647
\bibitem[Pinto et al.(2015)]{pine15} Pinto, R.F., Vilmer, N., Brun, A.S., 2015. Soft X-ray emission in kink-unstable coronal loops, Astron. Astrophys., 576, ID A37. http://doi.org/10.1051/0004-6361/201323358
\bibitem[Pinto et al.(2016)]{pine16} Pinto, R.F., Gordovskyy, M., Browning, P.K., Vilmer, N., 2016. Thermal and non-thermal emission from reconnecting twisted coronal loops, Astron. Astrophys., 585, ID A159. http://doi.org/10.1051/0004-6361/201526633
\bibitem[Pontin(2011)]{pont11} Pontin, D., 2011. Three-dimensional magnetic reconnection regimes: A review, Adv. Sp. Res., 47, 1508-1522. http://doi.org/10.1016/j.asr.2010.12.022
\bibitem[Pritchett(2008)]{prit08} Pritchett, P.L., 2008, Collisionless magnetic reconnection in an asymmetric current sheet, J. Geophys. Res., 113, ID A06210. http://doi.org/10.1029/2007JA012930
\bibitem[Pucci et al.(2017)]{puce17} Pucci, F., Servidio, S., Sorriso-Valvo, L., Olshevsky, V., Matthaeus, W.H., Malara, F., Goldman, M.V., Newman, D.L., Lapenta, G., 2017. Properties of Turbulence in the Reconnection Exhaust: Numerical Simulations Compared with Observations, Astrophys. J., 841, ID 60. http://doi.org/10.3847/1538-4357/aa704f
\bibitem[Rieke et al.(2015)]{riee15} Rieke, M., Trost, T., Grauer, R., 2015. Coupled Vlasov and two-fluid codes on GPUs, J. Comp. Phys. 283, 436-452. http://doi.org/10.1016/j.jcp.2014.12.016
\bibitem[Ripperda et al.(2017a)]{ripp17a} Ripperda, B., Porth, O., Xia, C., Keppens, R., 2017. Reconnection and particle acceleration in interacting flux ropes - I. Magnetohydrodynamics and test particles in 2.5D, Mon. Not. Roy. Astr. Soc, 467, 3279-3298. http://doi.org/10.1093/mnras/stx379
\bibitem[Ripperda et al.(2017b)]{ripp17b} Ripperda, B., Porth, O., Xia, C., Keppens, R., 2017. Reconnection and particle acceleration in interacting flux ropes - II. 3D effects on test particles in magnetically dominated plasmas, Mon. Not. Roy. Astr. Soc, 471, 3465-3482. http://doi.org/10.1093/mnras/stx1875
\bibitem[Rosdahl and Galsgaard(2010)]{roga10} Rosdahl, K.J, Galsgaard, K., 2010. Test particle acceleration in a numerical MHD experiment of an anemone jet, Astron. Astrophys., 511, ID A73. http://doi.org/10.1051/0004-6361/200913541
\bibitem[Saiz et al.(2013)]{saie13} Saiz, E., Cerrato, Y., Cid, C., Dobrica, V., Hejda, P., Nenovski, P., Stauning, P., Bochnicek, J., Danov, D., Demetrescu, C., Gonzalez, W.D., Maris, G., Teodosiev, D., Valach, F., 2013. Geomagnetic response to solar and interplanetary disturbances, J.Sp.Weath.Sp.Clim., 3, ID A26. http://doi.org/10.1051/swsc/2013048
\bibitem[Sakai(1990)]{saka90} Sakai, J.I., 1990. High-energy particle acceleration during the implosion driven by three-dimensional X-type current loop coalescence in solar flares, Astrophys.J., 365, 354-364. http://doi.org/10.1086/169490
\bibitem[Sato et al.(1982)]{sate82} Sato, T., Matsumoto, H., Nagai, K., 1982. Particle acceleration in time‐developing magnetic reconnection process, J.Geophys.Res., 87, 6089-6097.http://doi.org/10.1029/JA087iA08p06089
\bibitem[Schwenn(2006)]{schw06} Schwenn, R., 2006. Space Weather: The Solar Perspective, Liv.Rev.Solar Phys., 3, ID 2. http://doi.org/10.12942/lrsp-2006-2
\bibitem[Sharykin \& Kuznetsov(2016)]{shku16} Sharykin, I.N., Kuznetsov, A.A., 2016. Modelling of Nonthermal Microwave Emission from Twisted Magnetic Loops, Solar Phys., 291, 1341-1355. http://doi.org/10.1007/s11207-016-0917-y
\bibitem[Sharykin et al.(2018)]{shae18} Sharykin, I.N., Kuznetsov, A.A., Myshyakov, I.I., 2018. Probing Twisted Magnetic Field Using Microwave Observations in an M Class Solar Flare on 11 February, 2014, Solar Phys., 293, ID 34. http://doi.org/10.1007/s11207-017-1237-6
\bibitem[Shibata et al.(1995)]{shie95} Shibata, K., Masuda, S., Shimojo, M., et al., 1995. Hot-Plasma Ejections Associated with Compact-Loop Solar Flares, Astrophys. J. Lett., 451, L83-L85. http://doi.org/10.1086/309688
\bibitem[Siversky \& Zharkova(2009)]{sizh09} Siversky, T., Zharkova, V.V., 2009. Particle acceleration in a reconnecting current sheet: PIC simulation, J.Plasm.Phys., 75, 619-636. http://doi.org/10.1017/S0022377809008009
\bibitem[Snow et al.(2017)]{snoe17} Snow, B., Botha, G.J.J., R\'{e}gnier, S., Morton, R.J., Verwichte, E., Young, P.R., 2017. Observational Signatures of a Kink-unstable Coronal Flux Rope Using Hinode/EIS, Astrophys. J., 842, ID 16. http://doi.org/10.3847/1538-4357/aa6d0e
\bibitem[Somov et al.(2002)]{some02} Somov, B.V., Henoux, J.-C., Bogachev, S.A., 2002. Is it possible to accelerate ions in collapsing magnetic traps?, Adv. Sp. Res., 30, 55-60. http://doi.org/10.1016/S0273-1177(02)00179-5
\bibitem[Somov \& Kosugi(1997)]{soko97} Somov, B.V., Kosugi, T., 1997. Collisionless Reconnection and High-Energy Particle Acceleration in Solar Flares, Astrophys. J., 485, 859-868. http://doi.org/10.1086/304449
\bibitem[Speiser(1965)]{spei65} Speiser, T.W., 1965. Particle Trajectories in Model Current Sheets, 1, Analytical Solutions, J. Geophys. Res., 70, 4219-4226. http://doi.org/10.1029/JZ070i017p04219
\bibitem[Stanier et al.(2012)]{stbr12} Stanier, A., Browning, P.K., Dalla, S., 2012. Solar particle acceleration at reconnecting 3D null points, Astron. Astrophys., 542, ID A47. http://doi.org/10.1051/0004-6361/201218857
\bibitem[Stanier(2013)]{stan13} Stanier, A., 2013. Magnetic Reconnection and Particle Acceleration in Semi-Collisional Plasmas, Ph.D. thesis, University of Manchester, https://www.escholar.manchester.ac.uk/uk-ac-man-scw:211308 
\bibitem[Sugiyama \& Kusano(2007)]{suku07} Sugiyama, T., Kusano, K., 2007. Multi-scale plasma simulation by the interlocking of magnetohydrodynamic model and particle-in-cell kinetic model, J. Comp. Phys., 227, 1340-1352. http://doi.org/10.1016/j.jcp.2007.09.011 
\bibitem[Tajima(1987)]{taje87} Tajima, T., Sakai, J., Nakajima, H., Kosugi, T., Brunel, F., Kundu, M.R., 1987. Astrophys.J., 321, 1031-1048. http://doi.org/10.1086/165694
\bibitem[Tenbarge et al.(2014)]{tene14}  TenBarge,J. M., Daughton,W.,  Karimabadi, H.,  Howes, G.G. and Dorland,W., 2014. Collisionless reconnection in the large guide field regime: Gyrokinetic versus particle-in-cell simulations, Phys. Plas., 21, ID 020708. http://doi.org/10.1063/1.4867068
\bibitem[Threlfall et al.(2015)]{thre15} Threlfall, J., Neukirch, T., Parnell, C.E., Eradat Oskoui, S., 2015. Particle acceleration at a reconnecting magnetic separator, Astron. Astrophys., 574, ID A7. http://doi.org/10.1051/0004-6361/201424366
\bibitem[Threlfall et al.(2018)]{thre18} Threlfall, J., Hood, A.W., Browning, P.K., 2018. Flare particle acceleration in the interaction of twisted coronal flux ropes, Astron. Astrophys., 611, ID A40. http://doi.org/10.1051/0004-6361/201731915
\bibitem[Tomczak(1999)]{tomc99} Tomczak, M., 1999. YOHKOH observations of the Neupert effect, Astron. Astrophys., 342, 583-591. 
\bibitem[Tronci \& Camporeale(2015)]{trca15} Tronci, C., Camporeale, E., 2015. Neutral Vlasov kinetic theory of magnetized plasmas, Phys. Plasmas, 22, ID 020704. http://doi.org/10.1063/1.4907665
\bibitem[Tsiklauri(2011)]{tsik11} Tsiklauri, D., 2011. Particle acceleration by circularly and elliptically polarised dispersive Alfv\'{e}n waves in a transversely inhomogeneous plasma in the inertial and kinetic regimes, Phys. Plasm., 18, 092903-092919. http://doi.org/10.1063/1.3633531
\bibitem[Tsiklauri \& Haruki(2008)]{tsha08} Tsiklauri, D., Haruki, T., 2008. Physics of collisionless reconnection in a stressed X-point collapse, Phys. Plasmas, 15, 102902. http://doi.org/10.1063/1.2999532
\bibitem[Turkmani et al.(2005)]{ture05} Turkmani, R., Vlahos, L., Galsgaard, K., Cargill, P.J., Isliker, H., 2005. Particle Acceleration in Stressed Coronal Magnetic Fields, Astrophys. J. Lett., 620, L59-L62. http://doi.org/10.1086/428395
\bibitem[Turkmani et al.(2006)]{ture06} Turkmani, R., Cargill, P.J., Galsgaard, K., Vlahos, L., Isliker, H., 2006. Particle acceleration in stochastic current sheets in stressed coronal active regions, Astron. Astrophys., 449, 749-757. http://doi.org/10.1051/0004-6361:20053548
\bibitem[Vekstein and Browning(1997)]{vebr97} Vekstein, G.E., Browning, P.K., 1997. Electric-drift generated trajectories and particle acceleration in collisionless magnetic reconnection, Phys. Plas., 4, 2262-2268. http://doi.org/10.1063/1.872555
\bibitem[Vilmer(2012)]{vilm12} Vilmer, N., 2012. Solar flares and energetic particles, Phil.Trans.Roy.Soc., 370, 3241-3268. http://doi.org/10.1098/rsta.2012.0104
\bibitem[Vlahos et al.(2009)]{vlae09} Vlahos, L., Krucker, S., Cargill, P.S., 2009. The Solar Flare: A Strongly Turbulent Particle Accelerator, Lec. Notes Phys., 778, 157-221. http://doi.org/10.1007/978-3-642-00210-6\verb|_|5
\bibitem[Wood \& Neukirch(2005)]{wone05} Wood, P., Neukirch, T., 2005. Electron Acceleration in Reconnecting Current Sheets, Solar Phys., 226, 73-95. http://doi.org/10.1007/s11207-005-5686-y
\bibitem[Zharkova \& Gordovskyy(2004)]{zhgo04} Zharkova, V.V., Gordovskyy, M., 2004. Particle Acceleration Asymmetry in a Reconnecting Nonneutral Current Sheet, Astrophys. J., 604, 884-891. http://doi.org/10.1086/381966
\bibitem[Zharkova \& Gordovskyy(2005)]{zhgo05} Zharkova, V.V., Gordovskyy, M., 2005. Energy spectra of particles accelerated in a reconnecting current sheet with the guiding magnetic field, Mon. Not Roy. Astr. Soc., 356, 1107-1116. http://doi.org/10.1111/j.1365-2966.2004.08532.x
\bibitem[Zharkova et al.(2011)]{zhae11} Zharkova, V.V., Arzner, K., Benz, A.O., Browning, P., Dauphin, C., Emslie, A.G., Fletcher, L., Kontar, E.P., Mann, G., Onofri, M., Petrosian, V., Turkmani, R., Vilmer, N., Vlahos, L., 2011. Recent Advances in Understanding Particle Acceleration Processes in Solar Flares, Space Sci. Rev, 159, ID 357. http://doi.org/10.1007/s11214-011-9803-y
\bibitem[Zhou et al.(2015)]{zhoe15} Zhou, X., B{\"u}chner, J., B{\'a}rta, M., Gan, W., Liu, S., 2015. Electron Acceleration by Cascading Reconnection in the Solar Corona. I. Magnetic Gradient and Curvature Drift Effects, Astrophys. J., 815, ID 6. http://doi.org/10.1088/0004-637X/815/1/6
\bibitem[Zhou et al.(2016)]{zhoe16} Zhou, X., B{\"u}chner, J., B{\'a}rta, M., Gan, W., Liu, S., 2016. Electron Acceleration by Cascading Reconnection in the Solar Corona. II. Resistive Electric Field Effects, Astrophys. J., 827, ID 94. http://doi.org/10.3847/0004-637X/827/2/94
\end{thebibliography}
\end{document}